\documentclass[twocolumn,twoside,preprintnumbers,amsmath,amssymb,showkeys]{revtex4}

\usepackage{epsfig}

\usepackage{graphicx}

\usepackage{fancyhdr}
\usepackage{pslatex}

\begin{document}
\title {Selected results on Strong and Coulomb-induced correlations
from the STAR experiment}

\author{M. \v{S}umbera\footnote{e-mail: sumbera@ujf.cas.cz}
for the STAR Collaboration}

\affiliation{Nuclear Physics Institute, Academy of Sciences of the
Czech Republic, 250 68 \v{R}e\v{z}, Czech Republic}

\begin{abstract}
Using recent high-statistics STAR data from Au+Au and Cu+Cu
collisions at full RHIC energy I discuss strong and
Coulomb-induced final state interaction effects on identical
($\pi-\pi$) and non-identical ($\pi-\Xi$) particle correlations.
Analysis of $\pi-\Xi$ correlations reveals the strong and
Coulomb-induced FSI effects allowing for the first time to
estimate space extension of $\pi$ and $\Xi$ sources and average
shift between them. Source imaging technique providing clean
separation of these effects from effects due to the source
function itself is applied to one-dimensional relative momentum
correlation function of identical pions. For low momentum pions
and/or non-central collisions large departure from a
single-Gaussian shape is observed.

\end{abstract}


\maketitle

\thispagestyle{fancy}

\setcounter{page}{1}


\section{INTRODUCTION}
\vskip -0.4 cm

Progress in understanding space-time structure of multiparticle
production via femtoscopy technique is currently driven by
high-statistics data sets accumulated in heavy ion experiments at
RHIC and SPS accelerators \cite{LPSW05, WPCF05, Lednicky_WPCF05,
Lisa_WPCF05, Lednicky_QM05}. In particular ambitious program of
the STAR collaboration at RHIC exploiting good particle
identification has already provided vast variety of femtoscopic
measurements in different identical and non-identical particle
systems some of which were actually measured for the first time
\cite{Lisa_WPCF05}. This contribution is a progress report on two
currently pursued STAR femtoscopy analyses. Both of them were
introduced already at the previous WPCF meeting last year in
Krom\v{e}\v{r}\'{i}\v{z} \cite{WPCF05}. First focuses on
non-identical particle correlations in rather exotic meson--baryon
system $\pi^{\pm}-\Xi^{\pm}$ \cite{Chaloupka_WPCF05}. Previous
investigations have shown that correlations among these two
charged hadrons reveal not only the Coulomb but also the strong
final state interaction (FSI) effects
\cite{Chaloupka_QM05,Chaloupka_SQM06}. Order of magnitude
difference in mass plus  $\Delta$B=1/$\Delta$S=2 gap in baryon/
strangeness quantum numbers makes $\pi-\Xi$ system an important
tool to study the interplay between matter flow of on partonic and
hadronic level. The second analysis exploits particle correlations
in a more conventional system of two identical charged pions
\cite{Bystersky_WPCF05}. Its aim is to understand the geometry of
the source. Ultimate and rather ambitious aspiration of this
project is to extract pion-pion scattering lengths. This goal will
make heavy ion femtoscopic measurements fully competitive to a
dedicated particle physics experiments trying to extract this
important parameter of the strong interactions
\cite{Lednicky_WPCF05, Lednicky_CIPPQG01, Retiere_WPCF05}. Basic
prerequisites for such measurements are good knowledge of
correlations due to quantum statistics and Coulomb interactions.

\section{STUDYING SPACE-TIME STRUCTURE OF MULTI-STRANGE BARYON SOURCE
VIA $\pi-\Xi$ CORRELATIONS}

In ultra-relativistic heavy-ion collisions at RHIC hot and dense
strongly interacting matter is created exhibiting properties of
de-confined partonic matter \cite{STAR_white, QM05}. Almost
instantaneous equilibration of produced matter indicated by recent
heavy flavor measurements \cite{QM05, Abelev:2006db} represent one
of the greatest puzzles coming from the RHIC \cite{QM05}. The
early partonic collectivity also shows up in a subsequent
evolution of the system leading to strong collective expansion of
the bulk matter as demonstrated by large values of observed
elliptic flow \cite{STAR_white, QM05}. These observations are
further substantiated by STAR multi--strange baryon measurements
showing that $\Xi$ and $\Omega$ baryons reveal significant amount
of elliptic flow which is comparable to ordinary non--strange
baryons \cite{Adams:2005zg}. The sizable multi--strange baryon
elliptic flow which obeys constituent quark scaling
\cite{cq-scaling} confirms that substantial part of the collective
motion has indeed developed prior to hadronization. This picture
is also corroborated by more recent STAR analysis of the
centrality dependence of hyperon yields carried out within the
framework of a thermal model \cite{Adams:2006ke}. Observed scaling
behavior of strange baryons is consistent with a scenario of
hadron formation from constituent quark degrees of freedom through
quark recombination provided that the coalescence took place over
a volume that is much larger than the one created in any
elementary collisions.

These observations fit nicely into ideal hydro evolution starting
from the system of de-confined QCD matter. However they are also
consistent with more realistic hybrid macroscopic/microscopic
transport approach \cite{Nonaka:2006yn}  which takes into account
the strength of dissipative effects prevalent in the latter
hadronic phase of the reaction. The hybrid model calculations
indicate that at top RHIC energy the hadronic phase of the
heavy-ion reaction is of significant duration (at least 10 fm/c)
making hadronic freeze-out a continuous process, strongly
depending on hadron flavor and momenta. In particular heavy
hadrons, which are quite sensitive to radial flow effects, obtain
the additional collective "push" created by resonant
(quasi)elastic interactions during that fairly long-lived hadronic
rescattering stage \cite{Heinz:2004ar}.

It is clear that question concerning multi--strange baryon
decoupling from hot and strongly interacting partonic/hadronic
system is an interesting one but also not an easy one to solve.
Could this be provided by the femtoscopy? What kind of relevant
information can be obtained via low-relative-velocity correlations
of multi-strange baryons with the other hadrons? Since
non-identical particle correlations are sensitive not only to the
extent of the source, but also to the average shift in emission
time and position among different particle species \cite{LLED} the
answer is affirmative. The femtoscopy of non-identical particles
was already used to show that the average emission points of
pions, kaons and protons produced in heavy-ion collisions at SPS
and RHIC are not the same \cite{Lednicky_WPCF05,
Lednicky_CIPPQG01, Lednicky_QM05, Blume:2002mr, Adams:2003qa}. In
hydrodynamically inspired blast-wave approach
\cite{Retiere:2003kf} mass ordering of average space-time emission
points of different particle species naturally appears due to the
transverse expansion of the source. This effect increases with a
mass difference of the measured particle pair. Hence studying
correlations in the $\pi-\Xi$ system where the mass difference is
really big  should provide rather sensitive test of the emission
asymmetries introduced by the transverse expansion of the bulk
matter.

Moreover, in addition to the Coulomb interaction studied in the
$\pi-K$ system the small relative momentum $\pi^{\pm}-\Xi^{\mp}$
correlations may provide sufficiently clear signal of the strong
interaction revealing itself via $\Xi^*(1530)$ resonance.
Expressing particle momentum in the pair rest frame $\mathbf{k^*}
= \mathbf{p}_{\pi} = -\mathbf{p}_{\Xi}$ via pair invariant mass
$M_{\pi\Xi}$ and $m_{\pi}$ and $m_{\Xi}$

\vskip -.3cm
\begin{equation}
k^{*}= \frac{[M_{\pi\Xi}^2-(m_{\pi}-m_{\Xi})^2]^{1/2}
[M_{\pi\Xi}^2-(m_{\pi}+m_{\Xi})^2]^{1/2}}{2M_{\pi\Xi}}
\end{equation}

\noindent one expects the $\Xi^*(1530)$ peak to show up in the
correlation function $C(k^*)$ at $k^*\approx$ 150 MeV/c. Due to
its rather long lifetime $\tau_{\Xi^*(1530)} = 22 fm/c$ the
resonance could be also a rather sensitive to the $\pi-\Xi$
interaction during long-lived hadronic phase. This should be
investigated by both the $\pi-\Xi$ femtoscopy as well as via
direct measurements of the $\Xi^*(1530)$ spectra. While first
signal of the $\Xi^*$ resonance in heavy ion collision was seen in
femtoscopy analysis just two years ago \cite{Chaloupka:2004cf}
first preliminary STAR measurements concerning the $\Xi^*$ spectra
and their yields were presented only recently at the Quark Matter
conference in Shanghai this year \cite{RWitt_QM06}.

\subsection{Data selection}

Though in previous analyses \cite{Chaloupka:2004cf,
Chaloupka_QM05, Chaloupka_WPCF05, Chaloupka_SQM06} the $\pi-\Xi$
correlations were studied for two different system d+Au and Au+Au
and also at two different energies in this contribution I will
concentrate on Au+Au data at $\sqrt(s_{NN})$ = 200 GeV from RHIC
Run IV only. The data were divided into several centrality
classes. During the run central trigger was used to enhance
fraction of 10\% most central events. Track-level cuts based on
$dE/dx$ particle identification in the STAR Time Projection
Chamber were used. Pion sample momenta $p_t$ were limited to
[0.125, 1.] GeV/c. After the $dE/dx$ cuts the upper $p_T$-limit is
0.8GeV/c and 0.6 GeV/c at $y=0.$ and $y=0.8$, respectively.
Charged $\Xi$ were reconstructed topologically in the $p_t$ range
[0.7, 3.] GeV/c. To increase total number of analyzed $\pi-\Xi$
pairs in this analysis we have used wider rapidity cut then in the
previous STAR femtoscopy analyses \cite{Adams:2004yc}. The cut
$|y| < 0.8$ instead of $|y| < 0.5$ was employed for both particle
species. Total number of extracted $\Xi$ used in this analysis are
given in Table~\ref{tab1}.

\vskip -0.4cm
\begin{table}[htb]
\begin{center}
\caption{\bf 200GeV Au+Au, Run IV data set}
\begin{tabular}{|c|c|c|c|}
\hline
Centrality & No. of $\Xi^{\pm}$ & No. of $\Xi^{-}$
& No. of $\Xi^{+}$ \\
\hline
0 --10\% & $1084 \times 10^3$ & $595 \times 10^3$ & $489 \times 10^3$ \\
\hline
10 -- 40\% & $412 \times 10^3$ & $226 \times 10^3$ & $186 \times 10^3$ \\
\hline
40 -- 80\% &  $145 \times 10^3$ & $79 \times 10^3$ & $66 \times 10^3$ \\
\hline
\end{tabular}\label{tab1}
\end{center}
\end{table}

\subsection{Data analysis and corrections}

Event mixing technique was used to obtain uncorrelated
two-particle distribution in pair rest frame. To remove spurious
correlations of non-femtoscopic origin the uncorrelated pairs were
constructed from events with sufficient proximity in primary
vertex position along the beam direction, multiplicity and event
plane orientation variables. Pair cuts were used to remove effects
of track splitting and merging. Resulting raw correlation function
was then corrected for purity of both particle species. The
correction was performed individually for each bin in
$\mathbf{k^*} = (k^{*},cos\theta, \varphi)$ of the 3-dimensional
(3D) correlation function as described bellow.

\subsection{Pair purity analysis}

Pair purity defined as a fraction of primary $\pi-\Xi$ pairs is
calculated as a product of particle purities of both particle
species. $\Xi$ purity was obtained from reconstructed $\Xi$
invariant mass plot as a function of transverse momentum. Pion
purity was estimated using parameter $\sqrt{\lambda}$ of the
standard parametrization of the identical $\pi-\pi$ correlation
function. The identical pion measurements were performed with the
same pion cuts as those used in the $\pi-\Xi$ analysis. Since
value of the $\lambda$ parameter is influenced by decays of
long-lived resonance as well as by non-Gaussian shape of the
correlation function the pion purity correction can be a
significant source of the systematic error.

\begin{figure}[!htb1]
\begin{center}
\includegraphics*[angle=0., width=8 cm]{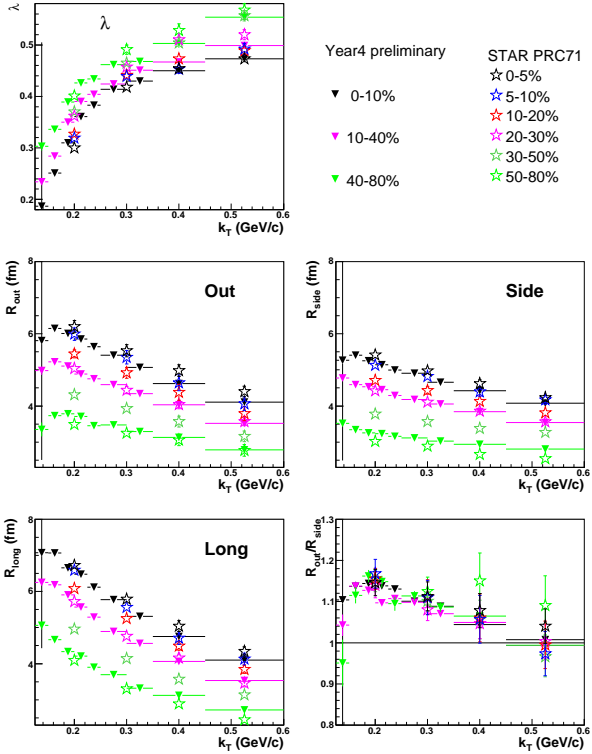}
\end{center}
\vskip -0.6cm \caption{\emph{ \small Comparison between parameters
of 3D Gaussian fit the correlation functions of identical charged
pions produced in Au+Au collisions at 200 GeV. Previous
\cite{Adams:2004yc} (open stars) and this analysis (full
triangles). Error bars contain only statistical uncertainties.}}
\end{figure}

In order to make contact with previous STAR identical pion
analyses \cite{Adams:2004yc} on Fig. 1  we present
$k_T$--dependence of parameters $\lambda$, $R_{out}$, $R_{side}$
and $R_{long}$ entering standard "out-side-long" decomposition of
3D correlation function $C(\mathbf{q})= 1 + \lambda \cdot
exp(-q_{out}^{2}R_{out}^{2} -q_{side}^{2}R_{side}^{2}
-q_{long}^{2}R_{long}^{2})$. Here $k_{T} = (|\mathbf{p}_{1T}| +
|\mathbf{p}_{1T}|)/2$ is average transverse momentum of two pions.
On the same figure the ratio $R_{out}/R_{side}$ is plotted too. We
conclude that the improved cuts used in the present analysis do
not change the values of extracted parameters but due to increased
acceptance in rapidity and transverse momenta of the pions our
analysis covers also region of lower $k_T$.

We have also studied influence of electrons on purity of the pion
sample. Exclusion of pions with $dE/dx$ within $\pm 2\sigma$
around the electron band has changed the value of parameter
$\lambda$ in the $k_T$ interval [0.15, 0.25] GeV/c by 50\% at
maximum. However the other parameters characterizing the 3D
correlation function of identical pions  (the radii
$R_{out}$,$R_{side}$,$R_{long}$) as well as the $\pi-\Xi$
correlation functions remained unchanged.

\subsection{Results in 1D – source size information}

The $\pi-\Xi$ correlation functions $C(k^{*})$ were analyzed in
the pair rest frame ($\mathbf{k}^{*} = \mathbf{p}_{\pi} =
-\mathbf{p}_{\Xi}$). The results for unlike sign pairs are
presented on Fig 2.

\begin{figure}[!htb1]
\begin{center}
\includegraphics*[angle=0., width=8 cm]{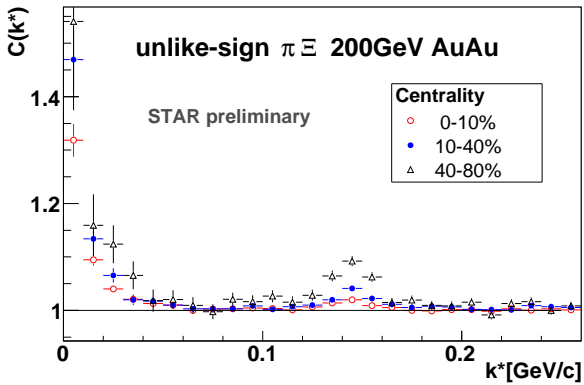}
\end{center}
\vskip -0.8cm \caption{\emph{ \small The centrality dependence of
the correlation function of combined sample of unlike-sign
$\pi^{\pm}–\Xi^{\mp}$ pairs .}}
\end{figure}

For all centralities the low $k^{*}$ region is dominated by the
Coulomb interaction. The strong interaction manifest itself in a
peak around $k^{*} \approx$ 150 MeV/c corresponding to
$\Xi^*$(1530). Peak's centrality dependence clearly shows high
sensitivity to the source size. Contrary to the Coulomb region the
correlation function in the resonance region does not suffer from
the low statistics and can thus in principle be used to extract
sizes of the sources more effectively then in the former case.

\subsection{Results in 3D – asymmetry measurement}

The information about shift in average emission points between
$\pi$ and $\Xi$ can be extracted from the angular part of the 3D
correlation function $C(\mathbf{k}^*)\equiv
C(k^{*},\cos\theta,\varphi)$.

\begin{figure}[!htb1]
\begin{center}
\includegraphics*[angle=0., width=8 cm]{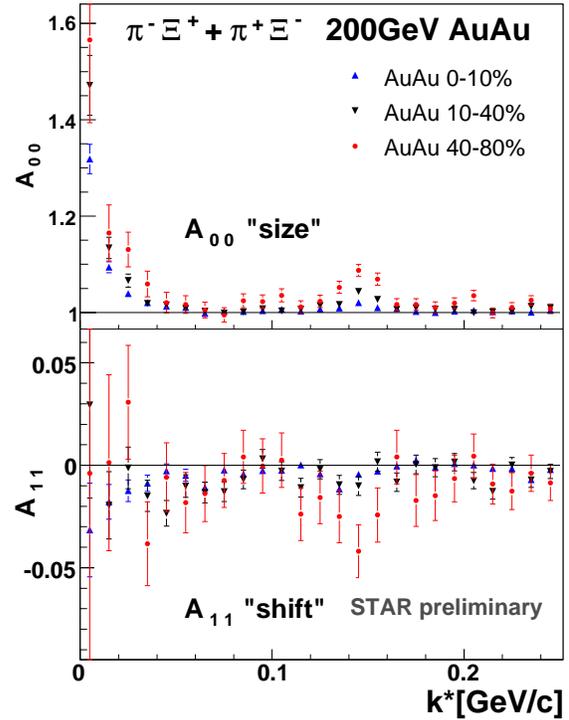}
\end{center}
\vskip -0.8cm \caption{\emph{ \small $A_{00}(k^*)$ and
$A_{11}(k^*)$ coefficients of spherical decomposition for combined
sample of unlike-sign $\pi^{\pm}–\Xi^{\mp}$ pairs from three
different centrality bins}}
\end{figure}

The function is binned in $k^*,\cos\theta,\varphi$  with
$\Delta_{\cos\theta}= \frac{2}{N_{cos\theta}}$ and
$\Delta_{\varphi}= \frac{2\pi}{N_{\varphi}}$ as bin sizes in
$\cos\theta$ and $\phi$, respectively. After its decomposition
into spherical harmonics \cite{Chajecki:2005}

\begin{equation}
A_{lm}(k^{*})=
\frac{\Delta_{\cos\theta}\Delta_{\varphi}}{\sqrt{4\pi}}\sum_{i}^{all~
bins}Y_{lm}(\theta_{i},\varphi_{i})
C(k^{*},\cos\theta_{i},\varphi_{i})
\end{equation}

\noindent symmetry constrains further limit number of relevant
components. This is due to the fact that individual coefficients
appearing in the above decomposition represent different
symmetries of the source. Thus for azimuthally symmetric identical
particle source at midrapidity, only $A_{lm}$ with even values of
$l$ and $m$ do not vanish. On the other hand for non--identical
particle correlations the coefficients with odd values of $l$ and
$m$ are allowed.

For both cases most important coefficient is $A_{00}(k^{*})$
representing angle-averaged correlation function $C(k^{*})$.
Latter is sensitive to the source size only. On the other hand in
non--identical particle case $A_{11}(k^{*})$ measures a shift of
the average emission point in the $R_{out}$ direction.

The decomposition of the correlation function into spherical
harmonics was used to study the centrality dependence of asymmetry
in emission between pions and $\Xi$. The results are presented on
Fig 3. The coefficient $A_{11}$ which is non-zero in all
centrality bins clearly confirms that the average space-time
emission points of pions and $\Xi$ are not the same.

\begin{figure}[!htb1]
\begin{center}
\includegraphics*[angle=0., width=8cm]{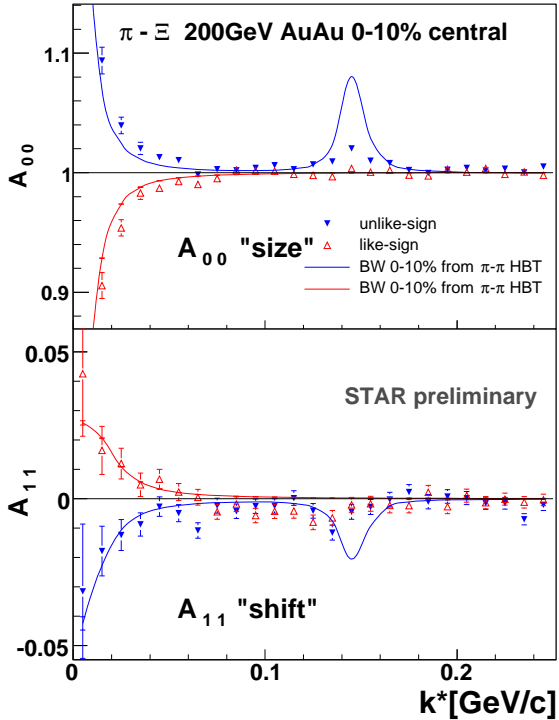}
\end{center}
\vskip -0.8cm \caption{\emph{ \small Comparison of $A_{00}(k^*)$
and $A_{11}(k^*)$ coefficients of spherical decomposition for
combined sample of unlike-sign $\pi^{\pm}–\Xi^{\mp}$ pairs from
10\% most central Au+Au collisions with the FSI model
predictions.}}
\end{figure}

On Fig. 4 the experimental results for the 10\%  most central, the
highest statistics bin are compared to a model calculation
exploiting strong and Coulomb final state interaction
\cite{Pratt:2003ar}.Theoretical correlation function was
constructed from particle emission separation distribution. For
this purpose momenta of real particles were used. The emission
coordinates of both $\pi$ and $\Xi$ were generated using the blast
wave model \cite{Retiere:2003kf}. This model encompassing
correlation between particle momenta and their space-time
coordinates was used with a single set of parameters  when
generating  emission coordinates of both $\pi$ and $\Xi$. These
parameters were obtained from experimentally measured pion spectra
and $\pi-\pi$ emission radii. Let us note that using the same set
of parameters for the $\Xi$ source as for the pions implicitly
assumes significant transverse flow of $\Xi$. In the Coulomb
region theoretical correlation function is in qualitative
agreement with the data. Moreover, orientation of the shift and
its magnitude agrees with the scenario in which $\Xi$ participates
in transverse expansion. However, in the region dominated by the
strong final state interaction the calculations over predict both,
the size and the shift coefficients.

\subsection{Extracting the source parameters}

For further analysis we have used only the low $k^*$ region
dominated by the Coulomb interaction, excluding thus the region
around the $\Xi^{*}(1530)$ peak (see Fig.5).

\begin{figure}[!htb1]
\begin{center}
\includegraphics*[angle=0., width=8cm]{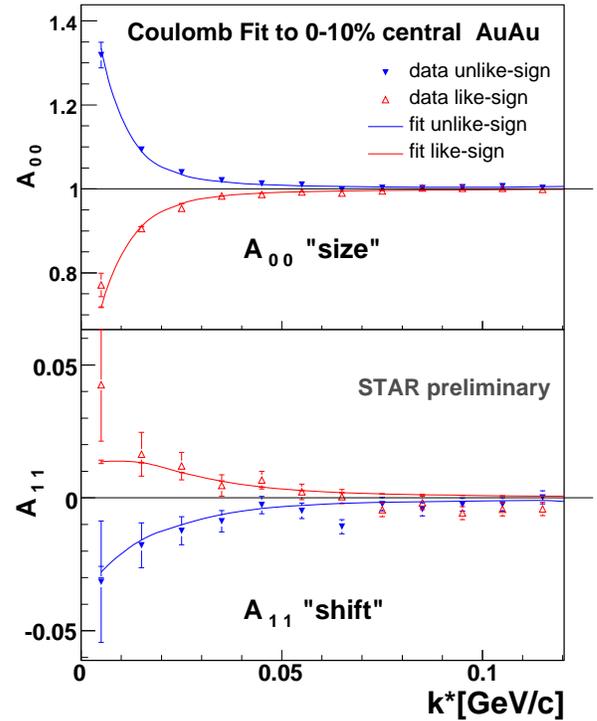}
\end{center}
\vskip -0.8cm \caption{\emph{ \small Comparison of $A_{00}(k^*)$
and $A_{11}(k^*)$ coefficients of spherical decomposition for
combined sample of like-sign $\pi^{\pm}–\Xi^{\pm}$ and unlike-sign
$\pi^{\pm}–\Xi^{\mp}$ pairs from 10\% most central Au+Au
collisions with the FSI model predictions.}}
\end{figure}

The theoretical correlation function was calculated using momentum
distribution of pairs extracted from the real data. The emission
coordinates were randomly generated from two-parameter source
distribution constructed in the following way. For both particle
species Gaussian shape of their source was assumed. The sources
were shifted in $R_{out}$ direction relative to each other. This
in the pair rest frame can be expressed via two parameters $R$ and
$\Delta_{out}$ characterizing the pair separation distribution
$\Delta r^*$. $R$ representing the width of the Gaussian and
$\Delta_{out}$ the shift in the $R_{out}$ direction. Values of the
source parameters were then extracted by finding a minimum value
of $\chi^2$ between theoretical and real correlation function.
Fitting was done simultaneously for like and unlike--sign
correlation functions.

For most central Au+Au collision this method yields first
preliminary values of $R=(6.7 \pm 1.0)$ fm and $\Delta_{out}=(-5.6
\pm 1.0)$ fm. The errors are purely statistical. Systematic error
studies are under way and their values are expected to be of the
order of the statistical ones. In our case a negative value of the
shift means that average emission point of $\Xi$ is positioned
more to the outside of the whole "fire-ball" than the average
emission point of pion as expected in the $\Xi$ flow scenario.

\section{IDENTICAL PION CORRELATIONS}

\subsection{Motivation}

In ultra--relativistic heavy--ion collisions femtoscopy can also
be employed to extract the information related to the strong
interaction among the particles \cite{Lednicky_WPCF05,
Lednicky_QM05}. In recent STAR analyses the strong FSI was studied
via two--$K^0_s$ interferometry \cite{Abelev:2006gu} and in
proton-lambda correlations \cite{Adams:2005ws}. A model that takes
the effect of the strong interaction into account has been used to
fit the measured correlation functions. In the $p -
\overline{\Lambda}$ and $\overline{p}- \Lambda$ correlations,
which were actually measured for the first time, annihilation
channels and/or a negative real part of the spin-averaged
scattering length was needed in the FSI calculation to reproduce
the measured correlation function.

At the previous WPCF meeting ambitious proposal to exploit the
correlations among identical charged pions to extract pion-pion
scattering lengths was made \cite{Retiere_WPCF05}. Potential for
such measurement at RHIC and latter on also at the LHC stems from
the fact that compared to a dedicated particle physics experiments
measuring scattering lengths parameters $a^0_0$ and $a^2_0$ like
BNL-E865 \cite{Pislak:2001bf} or Dirac \cite{Adeva:2005pg} heavy
ion experiments provide much larger number of pion pairs at small
relative momenta in a single event plus very large data samples (~
$10^7 - 10^9$ events). The real challenge when studying the strong
interaction among identical charged pions then is to beat down all
systematical errors pertinent to ultra--relativistic heavy ion
collisions environment. Coulomb interaction, pion purity and
geometry of the pion source need to be kept under control. Varying
source size ($k_T$ or centrality) may provide a good handle on
this.

In particular the bias arising from frequently used Gaussian
assumption of the source shape needs to be addressed. Using high
statistics sample of Au+Au events from STAR experiment at RHIC
highest energy accumulated during the run IV it was found that for
all $k_T$ and centrality bins the L\'{e}vy stable source
parametrization does not bring an advantage in describing the
detail shape of measured three-dimensional correlation function
\cite{Bystersky_WPCF05}.

Next step in this direction is to exploit model-independent
imaging technique of Brown and Danielewicz \cite{Brown:1997ku,
Brown:1997sn} and reconstruct the source itself. This is done via
inverting Koonin-Pratt equation (see e.g. \cite{LPSW05}):

\vskip -.3 cm
\begin{eqnarray}
C(q) - 1 = 4\pi \int{ K_{0}(q,r)S(r) r^2  dr}
\nonumber\\
K_{0}(q,r)\equiv \frac{1}{2}\int{(|\Phi(\mathbf{q},\mathbf{r})|^2
-1)d(cos\theta_{\mathbf{q},\mathbf{r}})}
\nonumber\\
q=\frac{1}{2}\sqrt{-(p_1 - p_2)^2}
\end{eqnarray}


\noindent which expresses the 1D correlation function $C(q)$ in
terms of the probability $S(r)$ to emit a pair of particles at a
separation $r$ in the c.m. frame. $S(r)$ is usually called source
function. All FSI is encoded in the (angle averaged) final state
wave function $\Phi(\mathbf{q},\mathbf{r})$ describing the
propagation of the pair from a relative separation of $\mathbf{r}$
in the pair c.m. to the detector with relative momentum
$\mathbf{q}$. The angle between vectors $\mathbf{q}$ and
$\mathbf{r}$ is denoted by $\theta_{\mathbf{q},\mathbf{r}}$. The
procedure to extract $S(r)$ via inversion of Eq. 3 is given in
\cite{Brown:1997sn}.

\begin{figure}[!htb2]
\includegraphics*[angle=0., width=7cm, height=11cm]{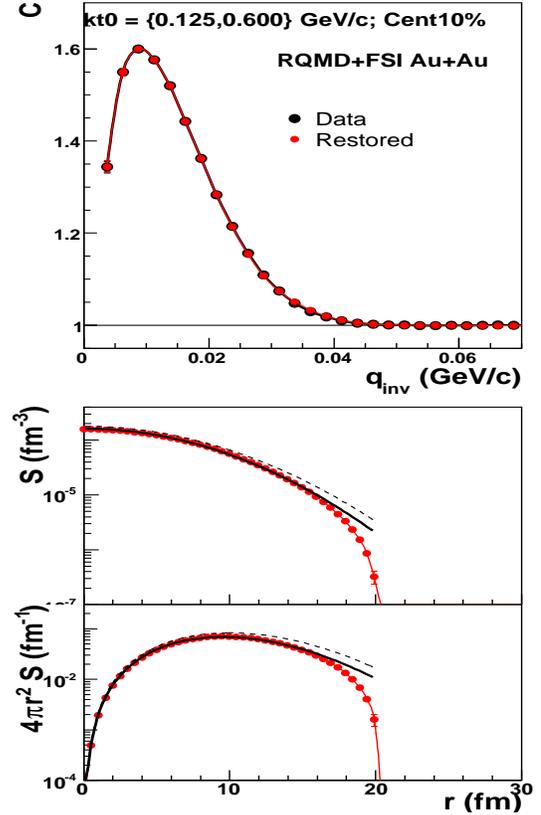}
\caption{\emph{\small 1D imaging analysis of isotropic Gaussian
source using pion momentum spectra from RQMD simulated 10\% most
central Au+Au collisions at $\sqrt{s_{NN}}$ =200GeV. Top: original
(black filled circles) and restored by imaging technique (red
filled circles) correlation function  $C(q)$. To guide an eye the
data points are connected by a smooth line of the same color.
Middle: 1D source function $S(r)$ (red filled circles). Full black
line represents result of a single-Gaussian fit to $S(r)$, dashed
line - input Gaussian distribution. Bottom: $4\pi r^2S(r)$.}}
\end{figure}

\subsection{Testing the inversion procedure}
To test reproducibility of original source function by the imaging
procedure we have generated Bose-Einstein (B-E) correlated and
Coulomb interacting pion pairs. The momentum spectra of the pions
were obtained from 10\% most central Au+Au RQMD generated events
at $\sqrt{s_{NN}}$ =200GeV. The pions coordinates were randomly
sampled from an isotropic 3D Gaussian distribution with the radius
R=5 fm. The instantaneous emission $\delta\tau$=0 of particles was
assumed. Using pion momenta and their coordinates the B-E
correlations and Coulomb FSI were introduced using procedure
described in \cite{coulomb}. Results for correlation and the
source functions are displayed on Fig.6. In order to compare with
the input Gaussian source the extracted source function was fitted
with Gaussian distribution again:

\vskip -.6cm

\begin{eqnarray}
g(r,R)= \frac{\lambda}{(2\sqrt{\pi}R)^{3}}{e^{-\frac{r^2}{4R^2}}}
\end{eqnarray}\label{SG}


Let us note that the $r^2$-weighting of $S(r)$ appearing in the
normalization condition of the source function:

\vskip -.6cm
\begin{equation}
4\pi\int_{0}^{\infty}S(r) r^2 dr = \lambda
\end{equation}

\noindent makes normalization constant $\lambda$ more sensitive to
behavior of $S(r)$ at $r \gg R$ than one would naively expect.
Since latter is determined by the imaging from values of the
correlation function $C(q)$ at very small $q$ fulfilment of the
above normalization condition using data with limited statistics
in the low $q$ bins may be hard to achieve. This may explain why
the values of extracted parameters $R$ and $\lambda$ from the
single-Gaussian fit were found to be smaller than the input ones.
While for $R$ the discrepancy is about 4\% for $\lambda$ it is
almost 25\%.

\begin{figure}[!htb2]
\resizebox{9pc}{!}{\includegraphics{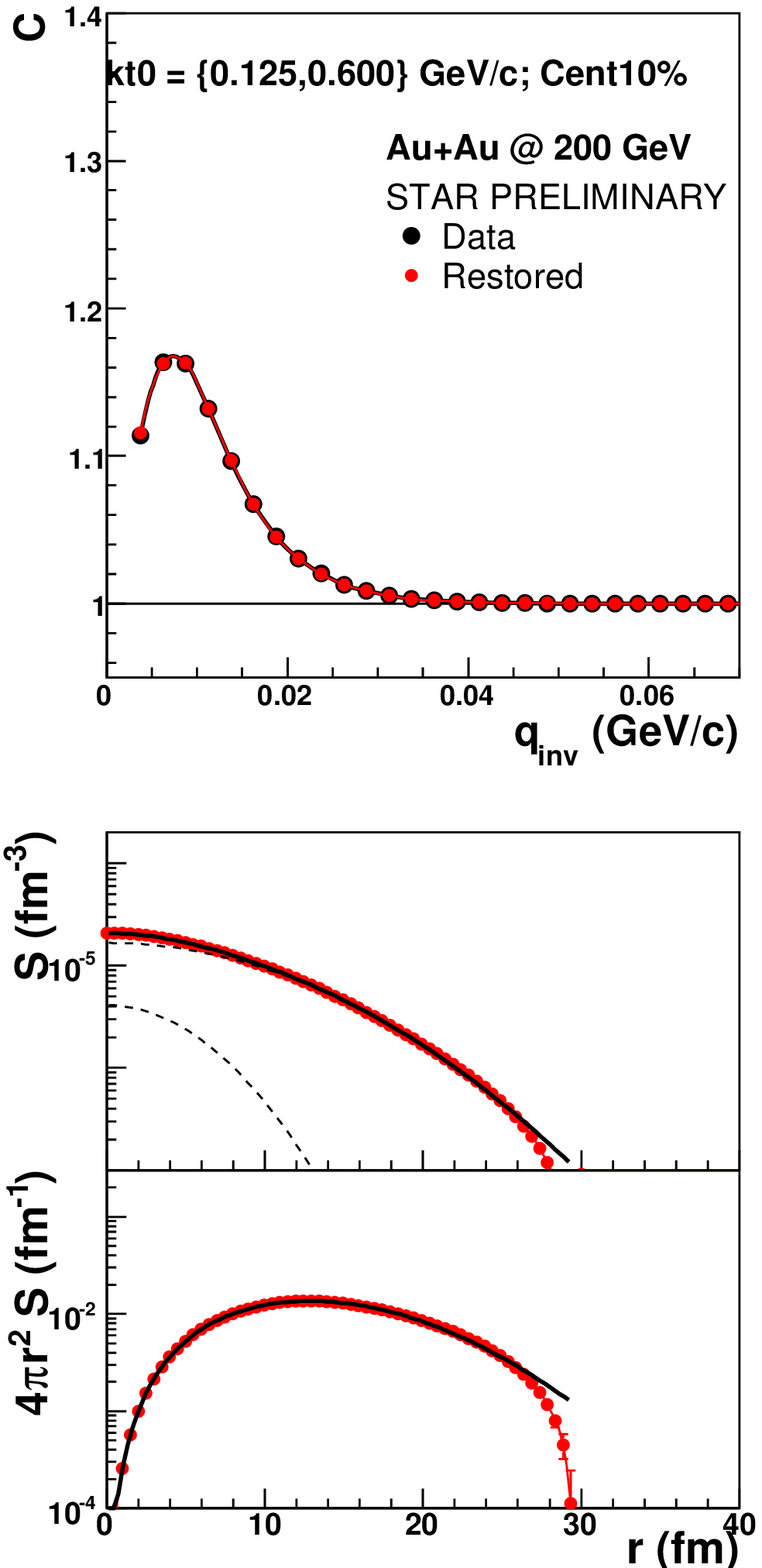}}
  \resizebox{9pc}{!}{\includegraphics{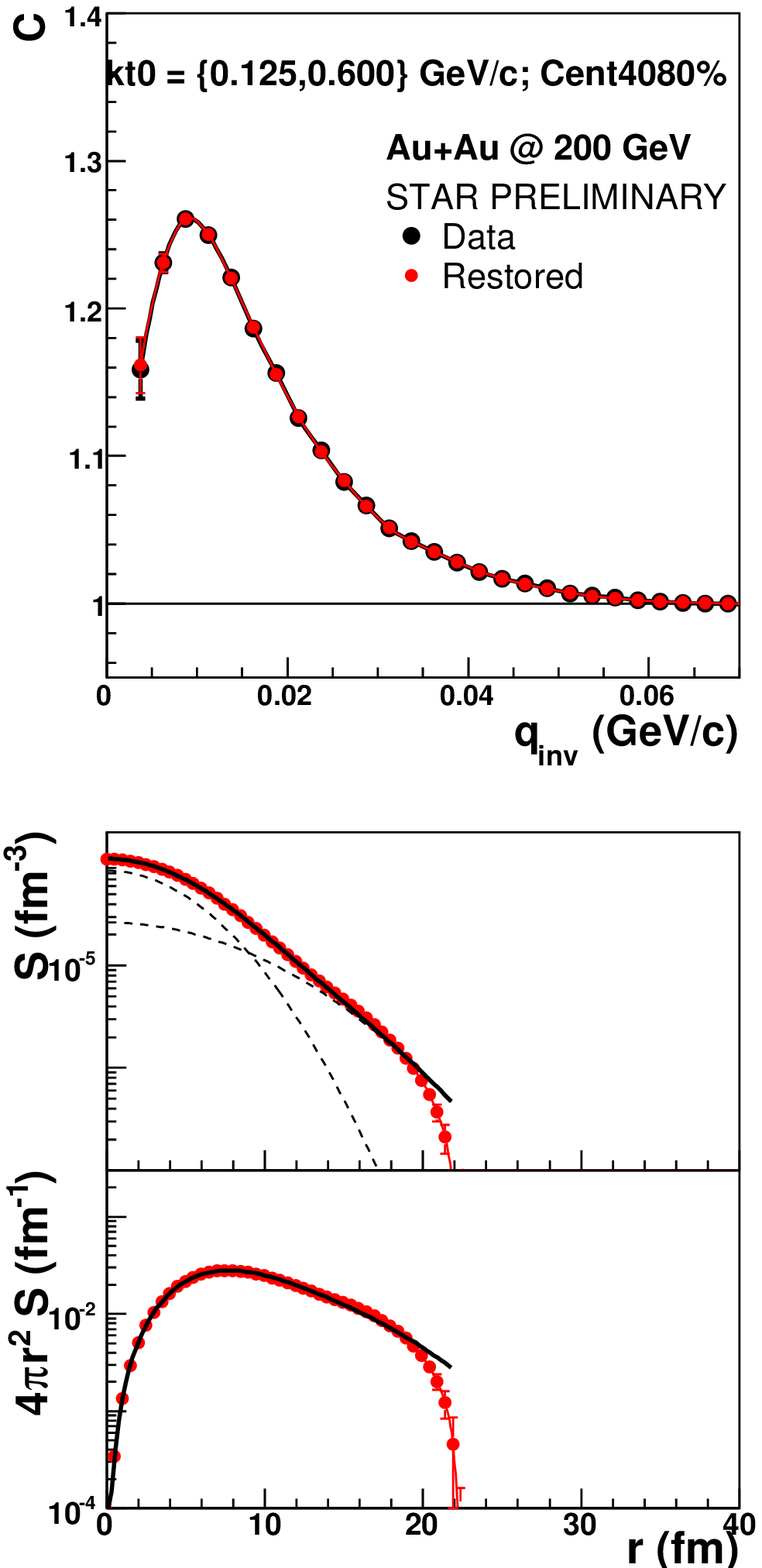}}\\
\caption{\emph{\small 1D imaging analysis of identical charged
pions from Au+Au collisions at $\sqrt{s_{NN}}$ =200GeV. Results
are shown for 10\% most central (the left panel) and  for
peripheral(centrality 40-80\%) collisions (the right panel). Top:
measured correlation function C(q) (black filled circles),
restored correlation function from imaging technique (red filled
circles). To guide an eye the data points are connected by a
smooth line of the same color. Middle: 1D source function $S(r)$
(red filled circles). Full black line represents result of a
double-Gaussian fit to $S(r)$. Two dashed lines show contribution
of each of two Gaussians. Bottom: $4\pi r^2S(r)$: (red filled
circles). Full black line represents result of a double-Gaussian
fit.}}
\end{figure}

\subsection{1D imaging analysis of the pion source
from Au+Au and Cu+Cu collisions}

The same data set of Au+Au events from Run IV used in already
described $\pi-\Xi$ correlation analysis was also used for the
reconstruction of the pion source. Second data set employed in
imaging analysis consists of about 8 million Cu+Cu minimum bias
events at $\sqrt{s_{NN}}$ =200GeV accumulated during the Run V.
Selected results from 1D imaging analysis of the pion source using
centrality selected Au+Au and Cu+Cu events are displayed on Fig. 7
and 8, respectively.

\begin{figure}[!htb2]
\resizebox{9pc}{!}{\includegraphics{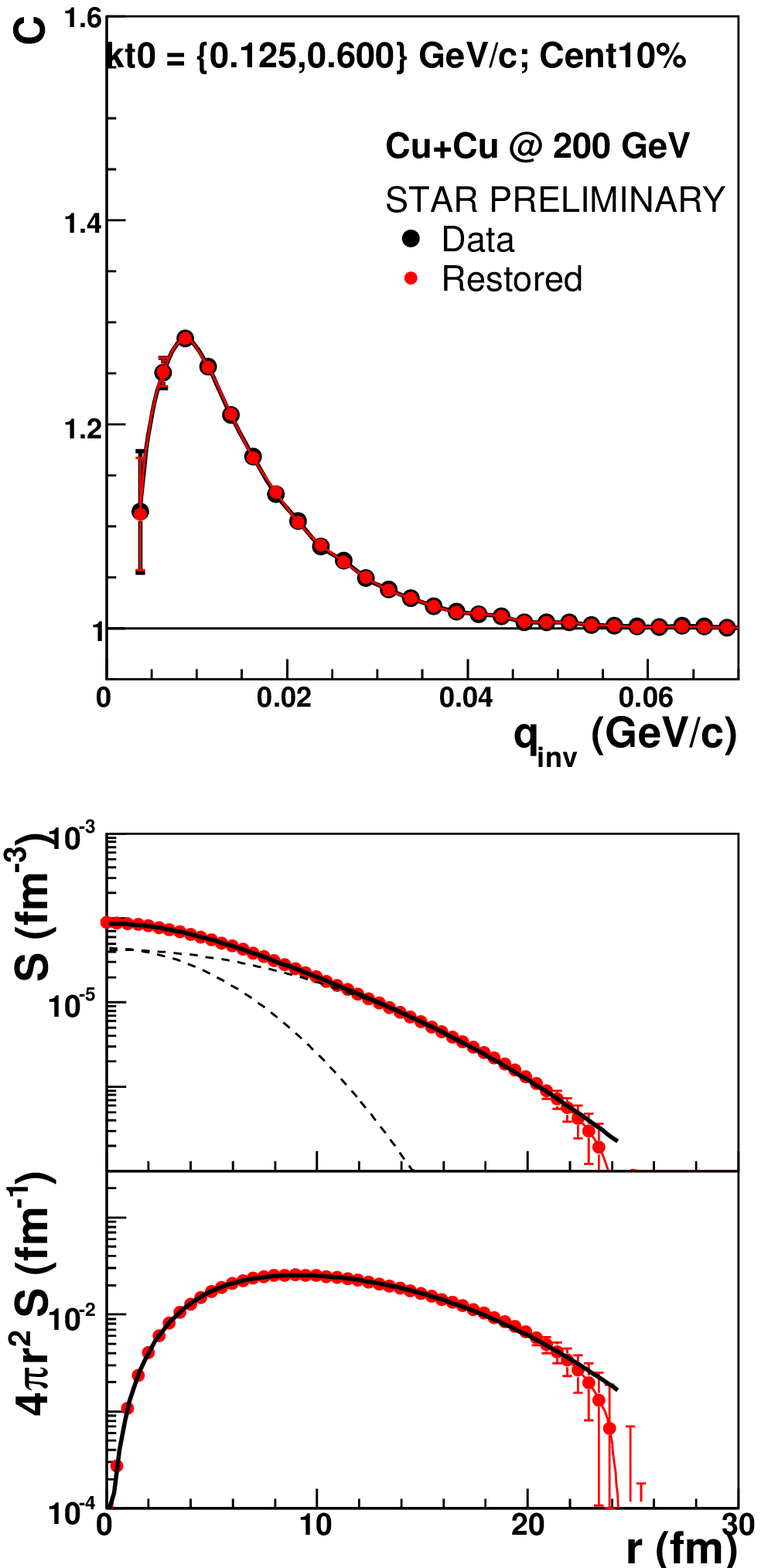}}
  \resizebox{9pc}{!}{\includegraphics{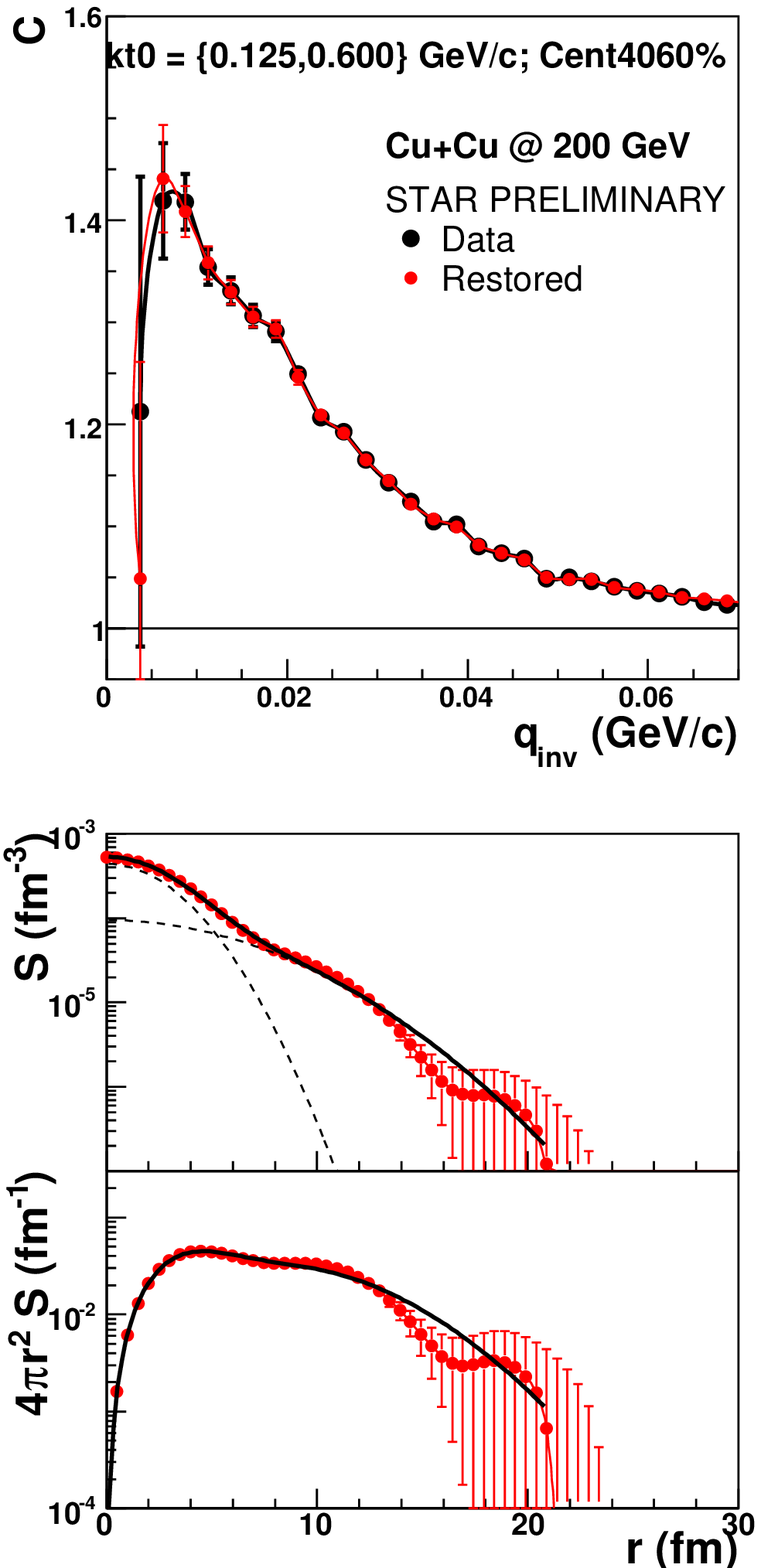}}\\
\caption{\emph{\small 1D imaging analysis of identical charged
pions from Cu+Cu collisions at $\sqrt{s_{NN}}$ =200GeV. Results
are shown for 10\% most central (the left panel) and  for
peripheral (centrality 40-60\%) collisions (the right panel). Top:
C(q). Middle: 1D source function $S(r)$. Bottom: $4\pi r^2S(r)$.
Labels are same as on Fig.7. }}
\end{figure}

Compared to model example discussed in the previous paragraph the
extracted source functions now develop long tails
\cite{Adler:2006as} and cannot thus be described by a single
Gaussian distribution. The only exception are the data from 10\%
most central Au+Au collisions (Fig.7.) To account for observed
behavior we use simplest extension of Eq.4 and assume that the
source function obtains contribution from two Gaussians. While the
first Gaussian $g(r,R_1)$ contributing with fraction $(1-\alpha)$
is responsible for the long tails the second one $g(r,R_2)$
contributing with weight $\alpha$ of width the $R_2 < R_1$
describes the source function at small $r$. The double Gaussian
distribution:

\vskip -.6cm

\begin{eqnarray}
G(r,R_1,R_2) = (1-\alpha)g(r,R_1) + \alpha g(r,R_2) , R_1
> R_2
\end{eqnarray}

\noindent now seems to describe data quite well. Comparing Au+Au
to Cu+Cu collisions of the same centrality we conclude that the
long tails are more pronounced for the smaller system.

All source functions presented so far were obtained from data
integrated over the whole range of average pair transverse momenta
$k_T$= [0.125, 0.600] GeV/c). In the following I will present
results showing how much observed departure from a single Gaussian
shape depends on particle transverse momenta.

\begin{figure}[!htb2]
\resizebox{9pc}{!}{\includegraphics{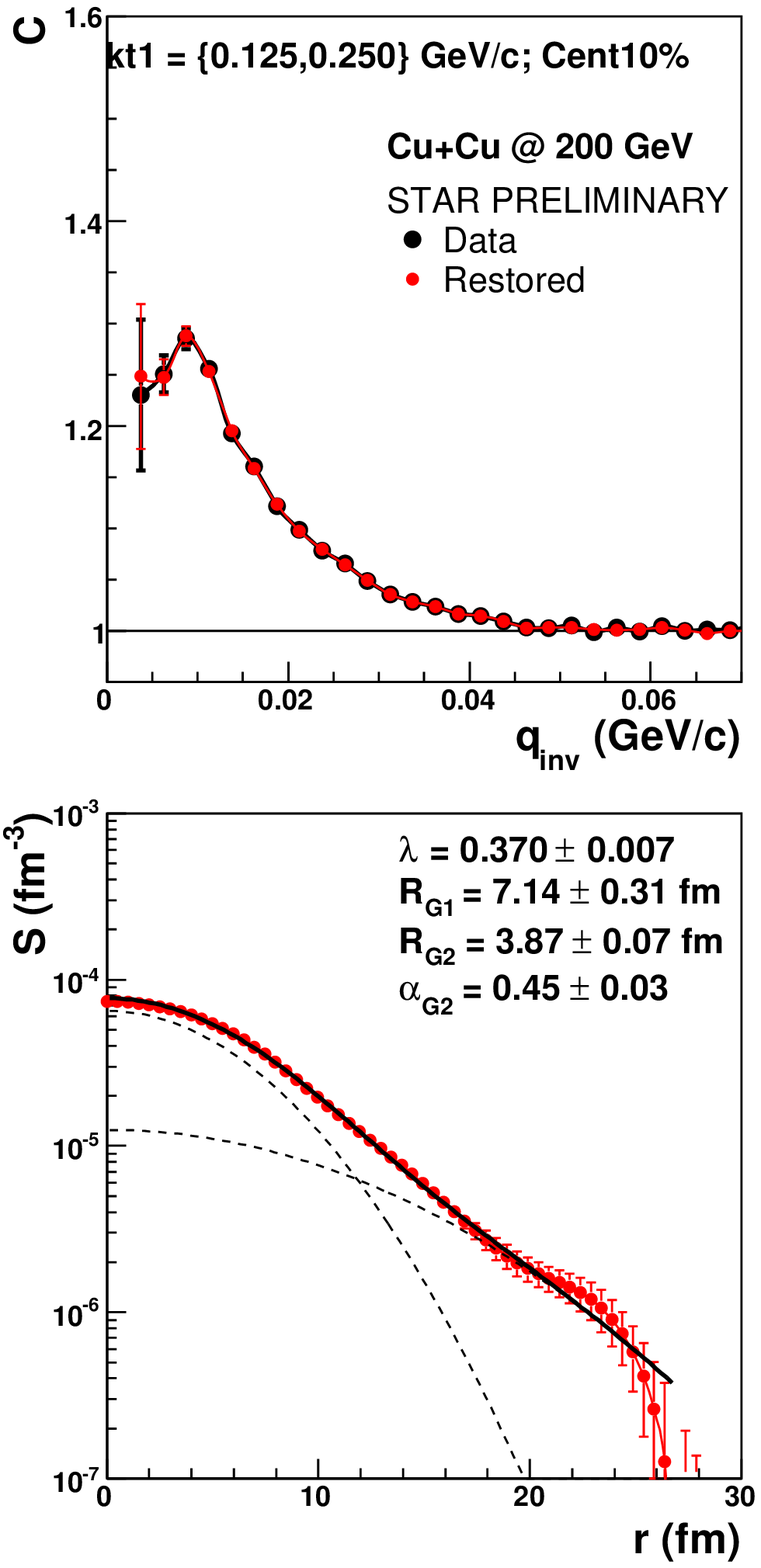}}
  \resizebox{9pc}{!}{\includegraphics{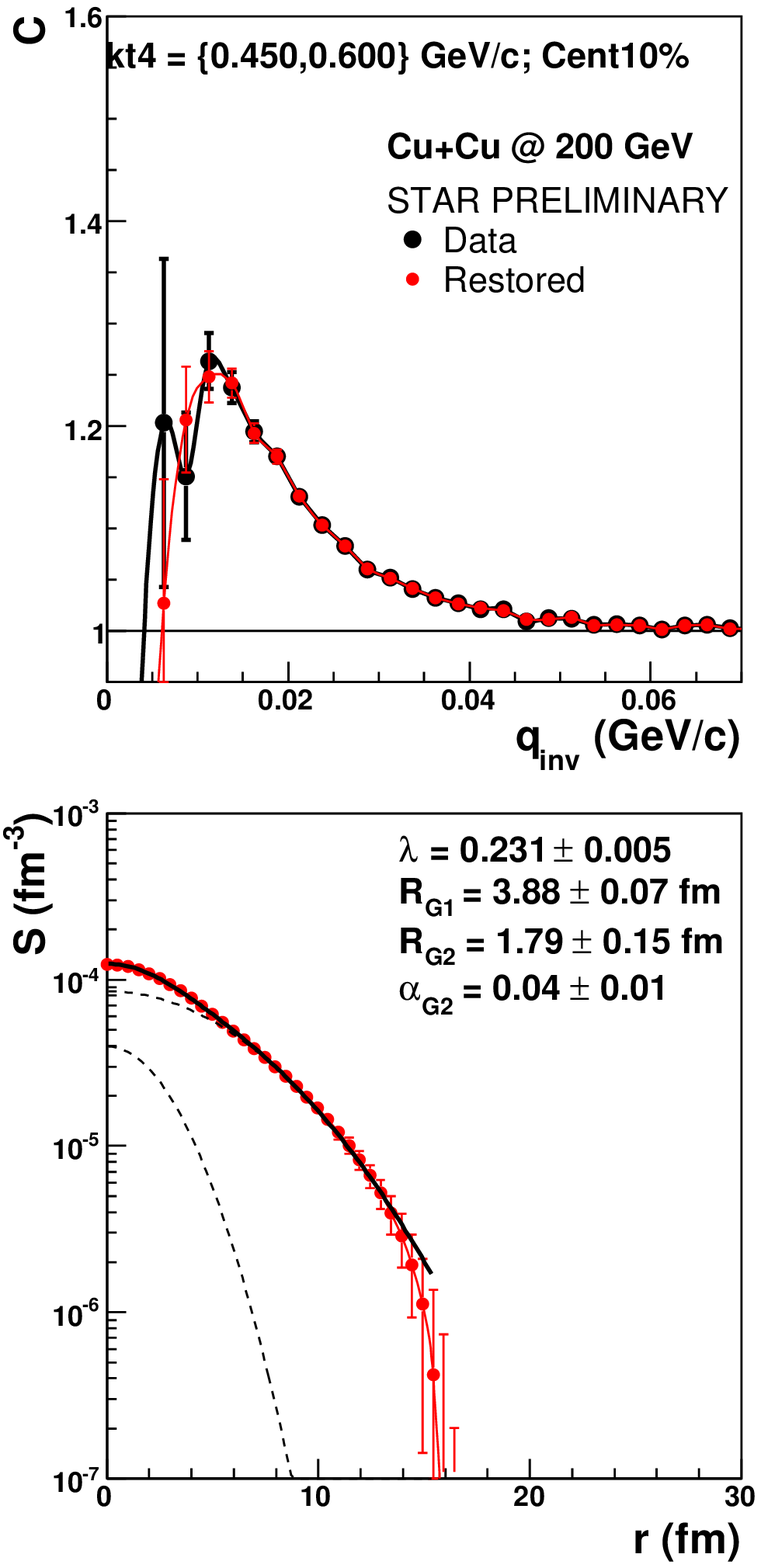}}\\
\caption{\emph{\small 1D imaging analysis of identical charged
pions from 10\%  most central Cu+Cu collisions at $\sqrt{s_{NN}}$
=200GeV. Results shown are for the lowest (the left panel) and for
the highest (the right panel) $k_T$-bin. Top:correlation function
C(q). Middle: 1D source function $S(r)$ Bottom: $4\pi r^2S(r)$.
Labels are same as on Fig.7.}}
\end{figure}

Such analysis is performed on Fig.9-11. While for Au+Au collisions
the statistics allowed us to use 9  bins in $k_T$ for Cu+Cu we
have used only 4 bins. On Fig.9 the lowest $k_T$= [0.125, 0.250]
GeV/c and the highest $k_T$= [0.450, 0.600] GeV/c bins from Cu+Cu
10\% most central collisions are compared. We see that the long
tail present in the lowest $k_T$ bin which is characterized by the
Gaussion with the width $R\approx$ 7 fm completely disappears from
the highest $k_T$ bin.

Fig. 10 and 11 show $k_T$-dependence of the double Gaussian fit
parameters for Cu+Cu and Au+Au collisions at three different
centralities, respectively. We observe that with increasing $k_T$
contribution of the second Gaussian is less and less prominent.
This behavior is more pronounced in Cu+Cu collisions.  Let us note
that long tails observed in the low-$k_T$ bins, then may indicate
an important r\^{o}le played by the pions from long-lived
resonance decays. An interesting observation which may have some
relevance to observed behavior regarding a random walk nature of
the particle freeze-out in coordinate space was made by
T.~Cs\"{o}rg\"{o} at this meeting \cite{Csanad:2007fr}.

\begin{figure}[!htb2]
\resizebox{21pc}{!}{\includegraphics{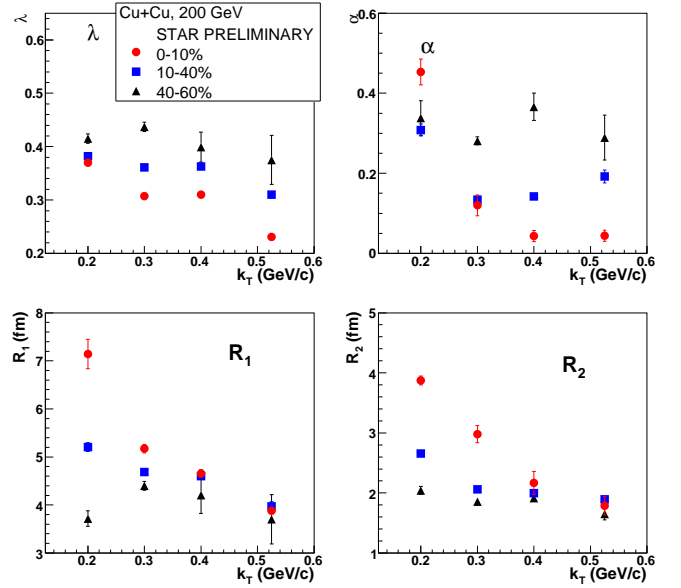}}%
\caption{\emph{\small $k_T$-dependence of the source parameters of
identical charged pions from Cu+Cu collisions at $\sqrt{s_{NN}}$
=200 GeV taken at three different centralities: 0-10\% - full red
circles, 10-40\% - full blue squares, 40-60\% - full black
triangles.}}
\end{figure}

\begin{figure}[!htb2]
\resizebox{21pc}{!}{\includegraphics{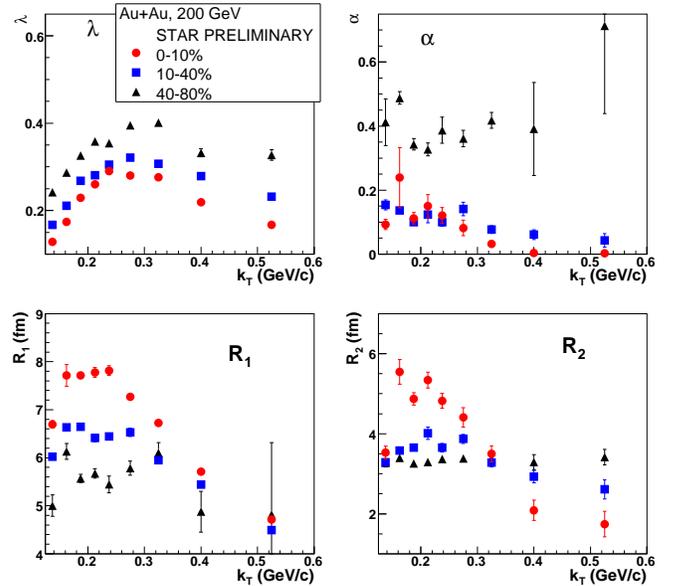}}%
\caption{\emph{\small $k_T$-dependence of the source parameters of
identical charged pions from Au+Au collisions at $\sqrt{s_{NN}}$
=200 GeV taken at three different centralities: 0-10\% - full red
circles, 10-40\% - full blue squares, 40-80\% - full black
triangles.}}
\end{figure}

\section{SUMMARY}
High-statistics STAR data from Au+Au and Cu+Cu collisions taken at
full RHIC energy and three different centralities were used to
discuss recent progress in identical ($\pi-\pi$) and non-identical
($\pi-\Xi$) particle femtoscopy. In $\pi-\Xi$ system the strong
and Coulomb-induced FSI effects are observed making it possible
for the first time to estimate the average shift and width between
$\pi$ and $\Xi$ source. 1D imaging of identical pion source
reveals significant departure from a single Gaussian shape.
Observed long tails which could be fairly well described by
allowing for the second Gaussian are more pronounced for the
sources produced at large impact parameters. The effect is
stronger for Cu+Cu then for Au+Au collisions. For all centralities
and both colliding systems the highest departure from single
Gaussian shape is observed for sources emitting particle with a
small average pair transverse momenta $k_T$. For highest analyzed
$k_T$ the source is to a good approximation a single Gaussian.

\break
\bigskip
\noindent\textbf{Acknowledgements}
\medskip

This work was supported in part by the IRP AV0Z10480505 and by
GACR grants 202/04/0793 and 202/07/0079. The data analysis was
performed by Michal Bystersk\'{y} and Petr Chaloupka as a part of
their Ph.D. thesis work. I would like to thank to both of them for
important input to my talk. My gratitude goes also to the
organizers of this excellent meeting, in particular to Sandra
Padula.

\end{document}